\documentclass[conference]{IEEEtran}
\usepackage{cite}
\usepackage{amsmath,amssymb,amsfonts}
\usepackage{graphicx}
\usepackage{array}
\usepackage{textcomp}
\usepackage{xspace}
\usepackage{color}
\usepackage{paralist}
\usepackage{wrapfig}
\usepackage{multirow}
\usepackage{listings}
\usepackage{makecell}
\usepackage{boxedminipage}
\usepackage{booktabs}
\usepackage{subcaption}
\usepackage{caption}
\usepackage{multirow}
\usepackage{multicol}
\usepackage{subcaption}
\usepackage{diagbox}
\usepackage{adjustbox}
\usepackage{url}
\usepackage{tabularx}

\usepackage{pifont}
\usepackage{textcomp}
\usepackage{balance} 
\usepackage{upquote}
\usepackage{amsthm} 
\usepackage[T1]{fontenc}
\usepackage{algorithm}
\usepackage{algpseudocode}

\usepackage{pifont}
\usepackage{balance}
\usepackage[htt]{hyphenat}
\usepackage{mdframed}
\usepackage{amsmath}
\newcommand{\coloneqq}{\mathrel{\mathop:}=}
\usepackage[table,xcdraw]{xcolor}
\usepackage[colorlinks=true,linkcolor=blue,citecolor=red,urlcolor=blue]{hyperref}

\usepackage[inline]{enumitem}

\usepackage{tcolorbox}

\tcbset{
    colback=gray!10, 
    colframe=black,  
    arc=3mm,         
    boxrule=0.5pt,   
    width=\linewidth-2mm, 
    left=3pt, right=3pt, top=3pt, bottom=3pt, 
}

\newcommand{\topone}[1]{\cellcolor{green!50}\textbf{#1}} 
\newcommand{\default}[1]{\cellcolor{green!10}#1} 

\newcommand{\refappendix}[1]{\hyperref[#1]{Appendix~\ref*{#1}}}

\newcounter{note}[section]

\newcommand{\ignore}[1]{}

\newcounter{packednmbr}

\newcounter{lessoncount}

\newcommand{\FB}{FLBuff\xspace}
\newcommand{\NIID}{non-iid\xspace}
\newcommand{\NIIDs}{non-iids\xspace}

\begin{document}

\title{Buffer is All You Need: Defending {F}ederated {L}earning against Backdoor Attacks under Non-iids via {Buff}ering}

\author{
    \IEEEauthorblockN{
          {Xingyu Lyu\IEEEauthorrefmark{1}},
          {Ning Wang\IEEEauthorrefmark{2}},
          {Yang Xiao\IEEEauthorrefmark{3}},
          {Shixiong Li\IEEEauthorrefmark{1}},
          {Tao Li\IEEEauthorrefmark{4}},
          {Danjue Chen\IEEEauthorrefmark{5}},
          {Yimin Chen\IEEEauthorrefmark{1}}
    }
    
    \IEEEauthorblockA{\IEEEauthorrefmark{1} Miner School of Computer and Information Sciences, University of Massachusetts Lowell, USA, \\ \IEEEauthorrefmark{2} Department of Computer Science and Engineering, University of South Florida, USA,\\ 
    \IEEEauthorrefmark{3}  Department of Computer Science, University of Kentucky,\\
    \IEEEauthorrefmark{4} Department of Computer and Information Technology, Purdue University, USA, \\ \IEEEauthorrefmark{5}Department of Civil, Construction, and Environmental Engineering, North Carolina State University, USA \\
    \IEEEauthorrefmark{1}\{xingyu\_lyu, shixiong\_li, ian\_chen\}@uml.edu, \IEEEauthorrefmark{2} ningw@usf.edu, \IEEEauthorrefmark{3}litao@purdue.edu, \IEEEauthorrefmark{4}dchen33@ncsu.edu
    }
}

\maketitle

\begin{abstract}
Federated Learning (FL) is a popular paradigm enabling clients to jointly train a global model without sharing raw data. However, FL is known to be vulnerable towards backdoor attacks due to its distributed nature. As participants, attackers can upload model updates that effectively compromise FL. What's worse, existing defenses are mostly \textit{designed} under independent-and-identically-distributed (iid) settings, hence neglecting the fundamental non-iid characteristic of FL. Here we propose \textit{FLBuff} for tackling backdoor attacks even under \NIIDs. The main challenge for such defenses is that \textit{non-iids bring benign and malicious updates closer, hence harder to separate.} \FB is inspired by our insight that \NIIDs can be modeled as omni-directional expansion in representation space while backdoor attacks as uni-directional. This leads to the key design of \FB, i.e., a supervised-contrastive-learning model extracting penultimate-layer representations to create a large in-between buffer layer. Comprehensive evaluations demonstrate that \FB consistently outperforms state-of-the-art defenses.

\end{abstract}	


\section{Introduction}
\label{sec:intro}

While Federated Learning (FL) has been widely recognized as one major framework for training machine learning (ML) models distributively, one main hurdle is its vulnerability towards poisoning attacks. The key aim of poisoning attacks is to either degrade the overall performance (i.e., untargeted attacks~\cite{fang2020local}) or manipulate the output predictions (i.e., targeted or backdoor attacks~\cite{bagdasaryan2019differential}) of the victim model. 
Among them, backdoor attacks on FL such as \textit{Scaling}~\cite{bagdasaryan2020backdoor} and \textit{ALIE}~\cite{baruch2019little} can use a simple inserted backdoor to trigger the victim model to output desired labels stealthily and effectively. 
Therefore, it is very critical to develop defense solutions for FL against them. 

In recent years, researchers have proposed defenses~\cite{cao2021fltrust,wang2022flare,zhangfldetecotr,zhang2023flipprovabledefenseframework,CrowdGuard,MESAS,fereidooni2023freqfed} that are shown to be effective. 
However, most of them are \textit{designed} under the assumption of independent-and-identically-distributed (iid) settings, neglecting the fundamental non-iid characteristic of FL. That is, the data distributions at different clients are inevitably different in practice, i.e., non-iid. Some defenses provide their non-iid evaluations however only under one or two types as illustrated in Table~\ref{table:flbuff-position}, which limits their applicability in real-world scenarios. 

\begin{table}[t]
    \centering
    \small
    \begin{tabular}{p{3.8cm}|p{4cm}} 
        \toprule
        Defense & \NIID categories evaluated \\ 
        \hline
        FLTrust~\cite{cao2021fltrust}, FLDetector~\cite{zhangfldetecotr},
        FLAME~\cite{nguyen2022flame}, FreqFed~\cite{fereidooni2023freqfed}   & prob \\
        \hline       FLIP~\cite{zhang2023flipprovabledefenseframework} & dir\\
        \hline
        FLARE~\cite{wang2022flare}  & qty\\
        \hline
        MESAS~\cite{MESAS}  & dir, qty \\
        \hline
        CrowdGuard~\cite{CrowdGuard} & prob, dir\\
        \hline
        \textbf{FLBuff}  & prob, dir, qty, noise, qs\\
        \bottomrule
    \end{tabular}
    \caption{A brief comparison between FLBuff and recent defenses with respect to \NIID categories evaluated.} 
    \label{table:flbuff-position}
    \vspace{-18pt}
\end{table}

Instead, we aim to propose a \textit{lightweight} and \textit{widely applicable} defense for FL against backdoor attacks, which works under a broad range of backdoors and non-iid scenarios. Essentially, we aim to bridge the above gap by taking a principled approach through answering three research questions. 

\begin{figure}[t]
    \centering
    \begin{subfigure}[b]{0.32\linewidth}
        \includegraphics[width=\linewidth]{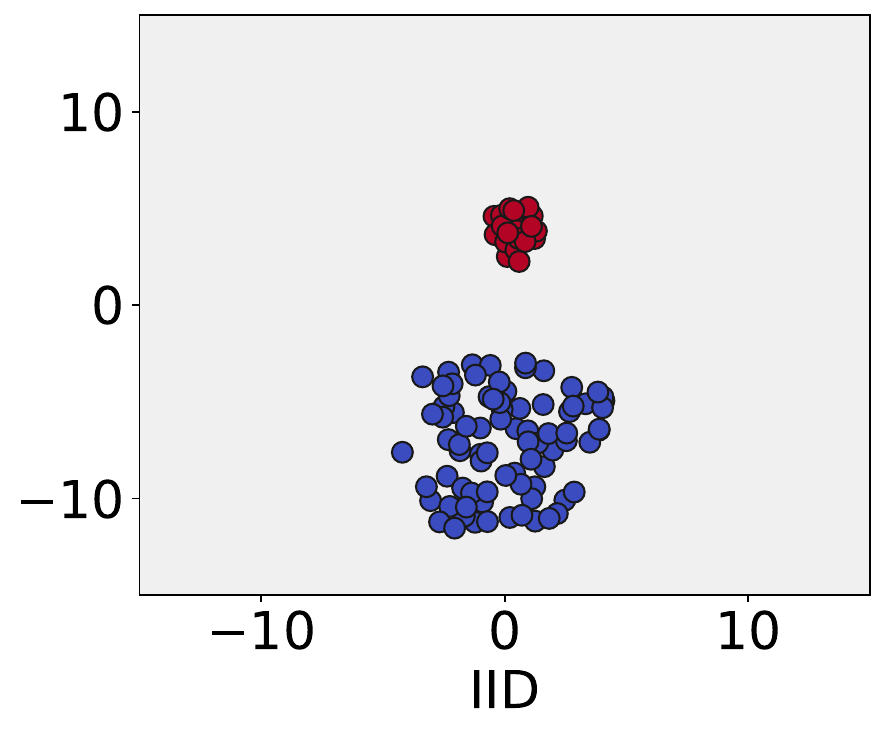}
    \end{subfigure}
    \hfill 
    \begin{subfigure}[b]{0.32\linewidth}
        \includegraphics[width=\linewidth]{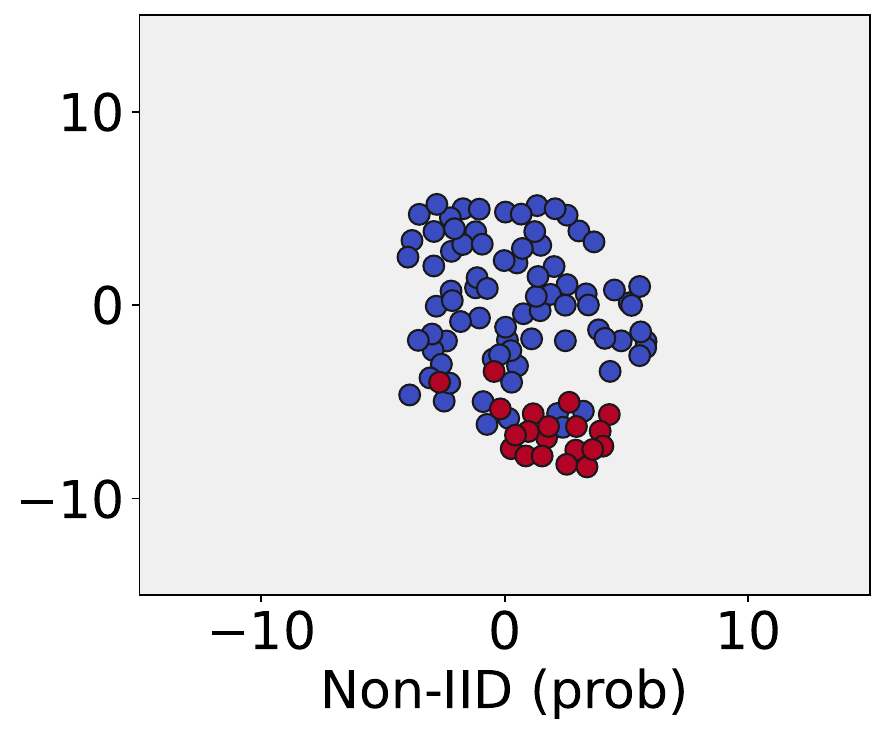}
    \end{subfigure}
    \hfill
    \begin{subfigure}[b]{0.32\linewidth}
        \includegraphics[width=\linewidth]{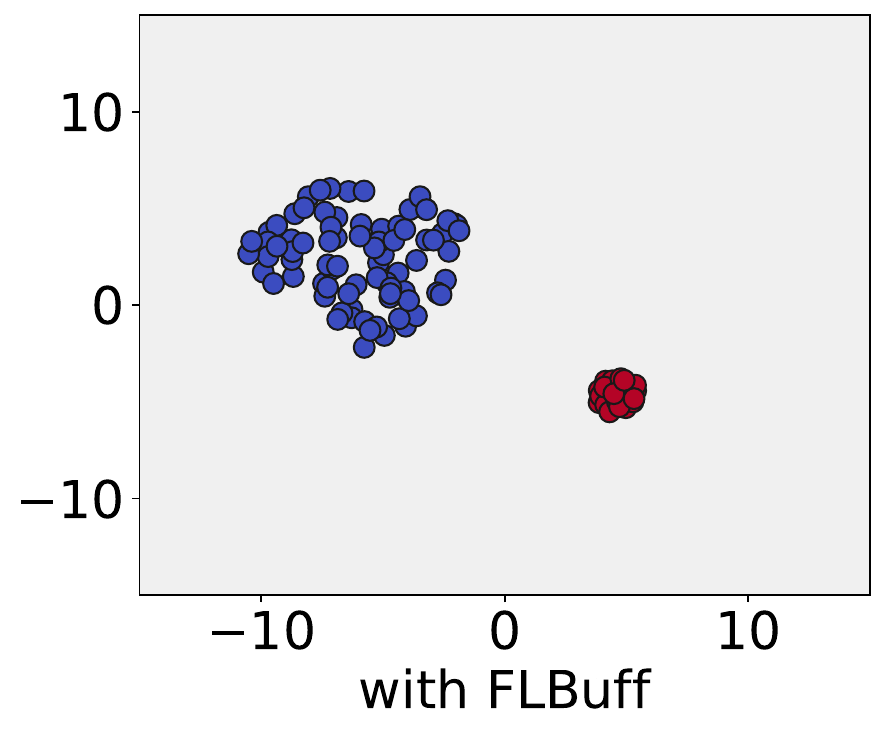}
    \end{subfigure}
    \begin{subfigure}[b]{\linewidth}
        \centering
        \includegraphics[width=0.8\linewidth]{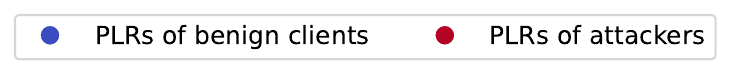} 
    \end{subfigure}
    \caption{Model updates for CIFAR-10 in PLR space under different scenarios: iid, \NIID (prob), and with \FB.}
    \label{fig:cl_fl}
    \vspace{-5pt}
\end{figure}

\textbf{RQ1:} What is the current landscape when it comes to defenses against backdoor attacks in FL under comprehensive \NIID settings? Our goal is to obtain key insights for why prior defenses would not work under various \NIIDs as well as provide a friendly benchmark framework for researchers to evaluate backdoor attacks and defenses under comprehensive \NIID settings. \textbf{RQ2:} Can we model the impacts of various \NIIDs and various backdoor attacks \textit{under the same perspective} so that by design, non-iids would not compromise our defense against backdoor attacks? Our investigations show that \NIIDs had significantly harmed prior defenses as benign model updates under \NIIDs became much less distinguishable from malicious ones (see Fig.~\ref{fig:cl_fl}). We aim to identify a representation space where the impacts of \NIIDs and backdoor attacks are differentiable. Following that, our third question emerges as \textbf{RQ3:} What is the key design under different \NIIDs so that the resulted solution is expected to work by design? 

\textbf{Contributions.}
We accomplish the following achievements in this paper. In answering \textbf{RQ1}, we develop an evaluation framework consisting of popular backdoor attacks and defenses under comprehensive \NIID settings including \textit{label-based, quantity-based, and feature-based \NIIDs}. The framework can be conveniently used by researchers to evaluate SOTA defenses on comprehensive attack profiles. 
In answering \textbf{RQ2}, we identify that from the perspective of representation space, \NIIDs and backdoor attacks are fundamentally different and therefore potentially differentiable. {On the one hand, the impact of \NIIDs can be modeled as omni-directional and small expansion of the cluster of benign model updates under iid. On the other hand, the impact of backdoor attacks resembles a large directional displacement from the same benign cluster, illustrated in Fig.~\ref{fig:small-buffer}. Prior defenses do not work well as \textit{the expansion from \NIIDs become overlapped with the cluster of backdoor model updates.} Hence our core design goal is \textit{to create a sufficiently large buffer layer} that can cover \NIID expansion and leave room between benign and backdoor clusters (see Fig.~\ref{fig:large-buffer}.)} Finally, we propose our solution for defending \underline{F}ederated \underline{L}earning against generic backdoor attacks under Non-IIDs via \underline{Buff}ering, i.e., \textbf{\FB}, featuring creating a viable in-between buffer layer by using a supervised-contrastive-learning-based (Sup-CL-based) loss, i.e., addressing \textbf{RQ3}. Particularly, \FB is shown to tackle a range of backdoor attacks under \NIIDs and even unknown ones by training from a labeled dataset under iid consisting of malicious model updates from only one backdoor attack and benign ones. We further summarize our contributions as follows.

\begin{itemize}
    \item We propose FLBuff, an effective supervised-learning-based defense against generic backdoor attacks under comprehensive \NIIDs. To the best of our knowledge, FLBuff is the first backdoor defense focusing on addressing \NIIDs in a systematic and comprehensive way.
    \item We identify the fundamental differences between backdoor attacks and \NIIDs based on an established unified view in representation space.
    \item We evaluate the performance of FLBuff through comprehensive experiments that encompass popular backdoor attacks and defenses across four datasets. Compared to five baselines, \FB consistently outperforms them. {Furthermore, our results also show that \FB can be applied to unseen backdoor attacks and is resilient towards various factors including strong adaptive attacks even under \NIIDs, hence confirming its practicality.}
\end{itemize}

\section{System and Adversary Model}
\label{sec:system_adversary_model}
\subsection{System Model}
\label{sec:system-model}
\label{sec:noniid}
\begin{figure}
    \centering
    \includegraphics[width=0.96\linewidth]{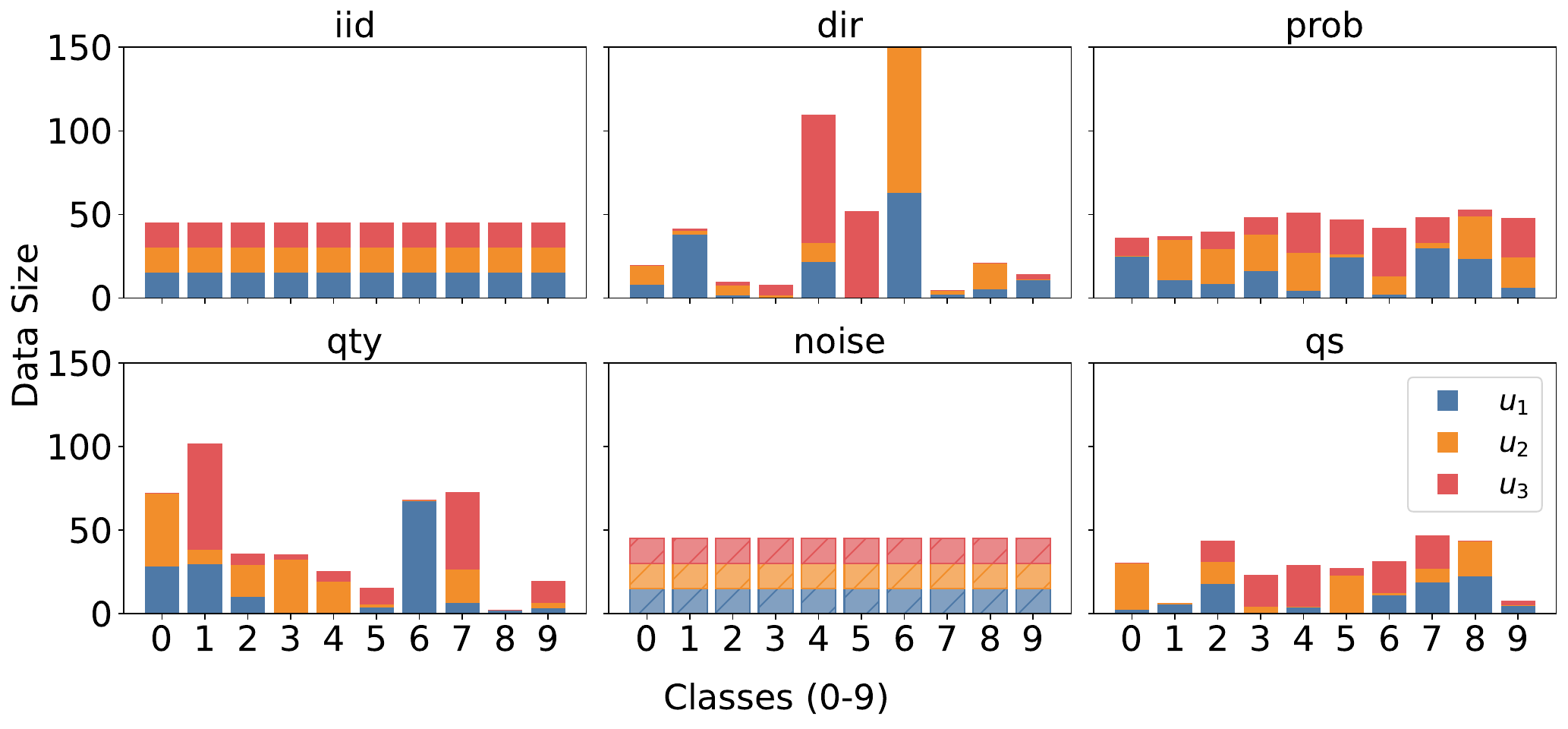}
    \vspace{-5pt}
   \caption{Illustrations of iid and five non-iids (dir-based, prob-based, qty-based, noise, qs) for $u_1, u_2$, and $u_3$, and 10 label classes.}
    \label{fig:iid-non-iid-examples}
    \vspace{-5pt}
\end{figure}

Our FL system consists of two main entities: a parameter server (PS) and $n$ remote clients (denoted as $[n] \coloneqq \{1, 2, \dots, n\}$). As discussed in Sect.~\ref{sec:intro}, we focus specifically on \NIID settings, where the local dataset $\mathcal{D}_i$ of a remote client $u_i$ follows certain non-identical distributions, i.e., non-iids. Fig.~\ref{fig:iid-non-iid-examples} provides illustrations of iid and the five \NIIDs we focus on in this paper, assuming a simple scenario with three clients ($u_1$, $u_2$, and $u_3$), 10 data classes (from label `0' to label `9'), and 150 samples per client. Here we only introduce them briefly while retaining the details in \textbf{supplementary}. First, iid refers to the case where different clients have the same number of data classes and the same number of samples per class. Second, we define the five investigated \NIID types briefly as below.



\begin{itemize}
    \item \textbf{Type-I \NIID (dir)}: Dirichlet distribution-based label imbalance, i.e., dir-based, where we allocate each client ($u_1, u_2$, and $u_3$) a certain number of samples per label (can even be 0) according to a given Dirichlet distribution. Existing defenses in~\cite{zhang2023flipprovabledefenseframework,CrowdGuard,MESAS} have explored this \NIID.
    \item \textbf{Type-II \NIID (prob)}: Probability-based label imbalance, i.e., prob-based, where clients ($u_1, u_2$, and $u_3$) are first divided into $G$ groups based on the main label class(es) they have (`0', `1', and `6' for $u_1$, $u_2$, and $u_3$, respectively) and then a given data sample is assigned to its primary group with probability $q$ and to other groups with probability $\frac{1-q}{G-1}$ as in \cite{fang2020local,cao2021fltrust,zhangfldetecotr,CrowdGuard}.
    \item \textbf{Type-III \NIID (qty) }: Quantity-based label imbalance, i.e., qty-based, where each client only has training samples from the same number of labels~\cite{wang2022flare,mcmahan2017communication}.
    \item \textbf{Type-IV \NIID (noise)}: Noise-based feature distribution skew where we add Gaussian noise to the samples of each client $u_i$ to achieve feature diversity~\cite{li2022federated} after dividing the dataset equally among $u_1, u_2$, and $u_3$.
    
    \item \textbf{Type-V \NIID (quantity-skew, qs)}: quantity skew where we use a Dirichlet distribution to allocate different number of samples to each client as in~\cite{li2022federated,luo2021no}. 
\end{itemize}
In brief, we find that rarely an existing defense against backdoor attacks had considered more than two \NIIDs while \FB is the first to tackle five most frequent ones.

\begin{figure}
    \centering
    \includegraphics[width=0.97\linewidth]{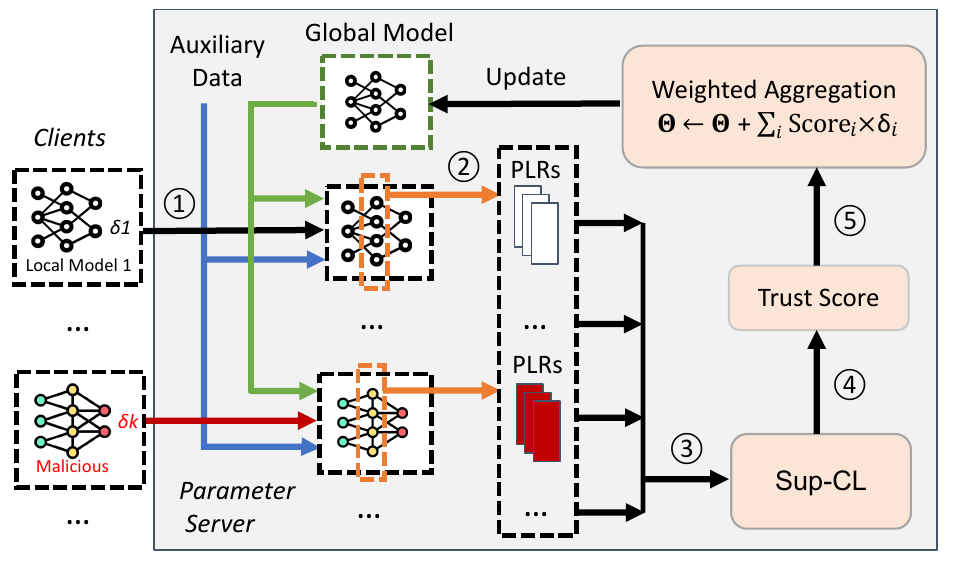}
    \vspace{-5pt}
    \caption{Workflow of FL with \textbf{\FB}.}
    \label{fig:flbuff-flow}
    \vspace{-5pt}
\end{figure}

\subsection{Adversary Model}


\textbf{Attackers' goals.} We adopt the attack goals from recent backdoor attacks~\cite{xie2019dba,bagdasaryan2020backdoor,baruch2019little}. On the one hand, an backdoor attacker aims to insert their backdoors into the global model $\Theta$ so that $\Theta$ is misled to classify specific input samples into the attacker's target label. On the other hand, $\Theta$ under attack needs to maintain a relatively high model accuracy on `clean' inputs so that the attack remains difficult to notice (i.e., stealthy). 

\textbf{Attackers' capabilities.} Similar to~\cite{xie2019dba,bagdasaryan2020backdoor,baruch2019little}, we assume that our backdoor attackers have access to a dataset that they can use to train their local models. Furthermore, they can arbitrarily manipulate (modify, add, or delete) their data, which can involve inserting triggers into samples and altering labels. They can manipulate their model updates as they want to achieve the above goals. We also assume backdoor attackers are treated as legitimate clients and therefore they have white-box access to $\Theta$ but no local models of other clients. We assume the number of malicious clients to be less than half of all clients. PS is trusted and to deploy \FB to properly protect $\Theta$. 
\section{Design of FLBuff}
\label{sec:method}

\begin{figure*}[ht]
    \centering
    \begin{subfigure}[b]{0.27\textwidth}
        \includegraphics[width=\textwidth]{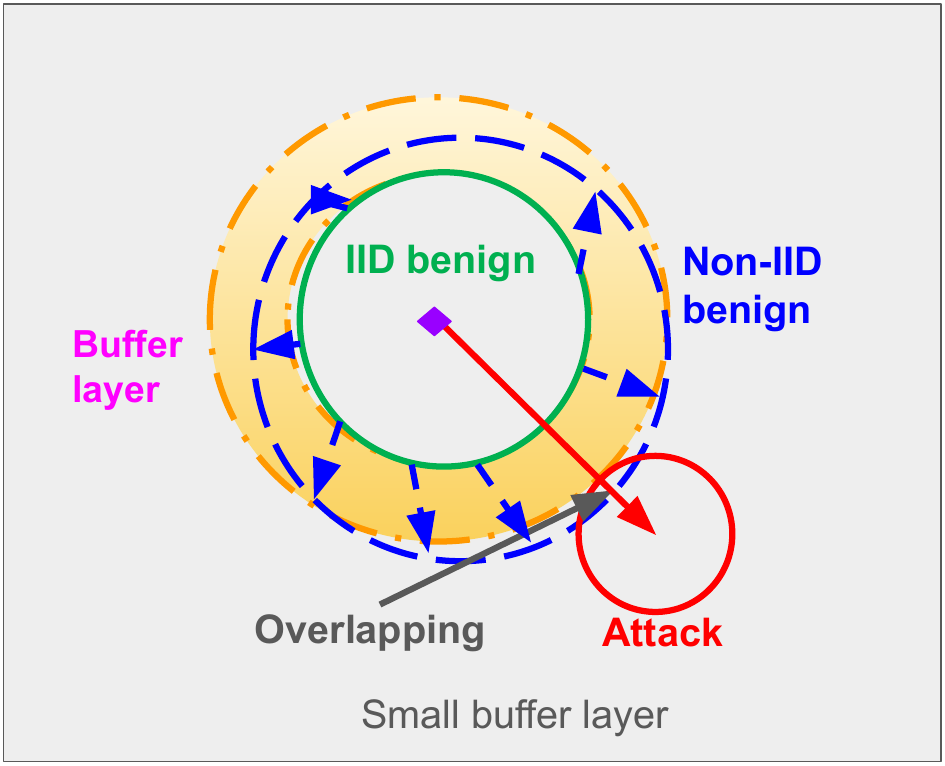}
        \caption{With small buffer layer}
        \label{fig:small-buffer}
    \end{subfigure}
    \begin{subfigure}[b]{0.27\textwidth}
        \includegraphics[width=\textwidth]{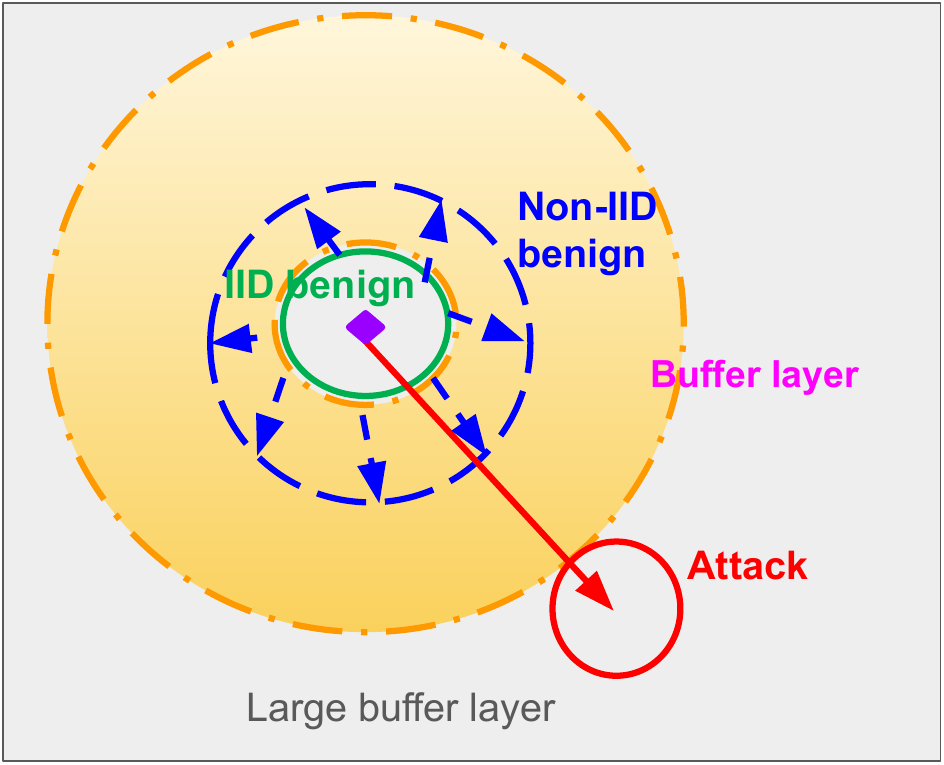}
        \caption{With large buffer layer}
        \label{fig:large-buffer}
    \end{subfigure}
    \begin{subfigure}[b]{0.27\textwidth}
        \includegraphics[width=\textwidth]{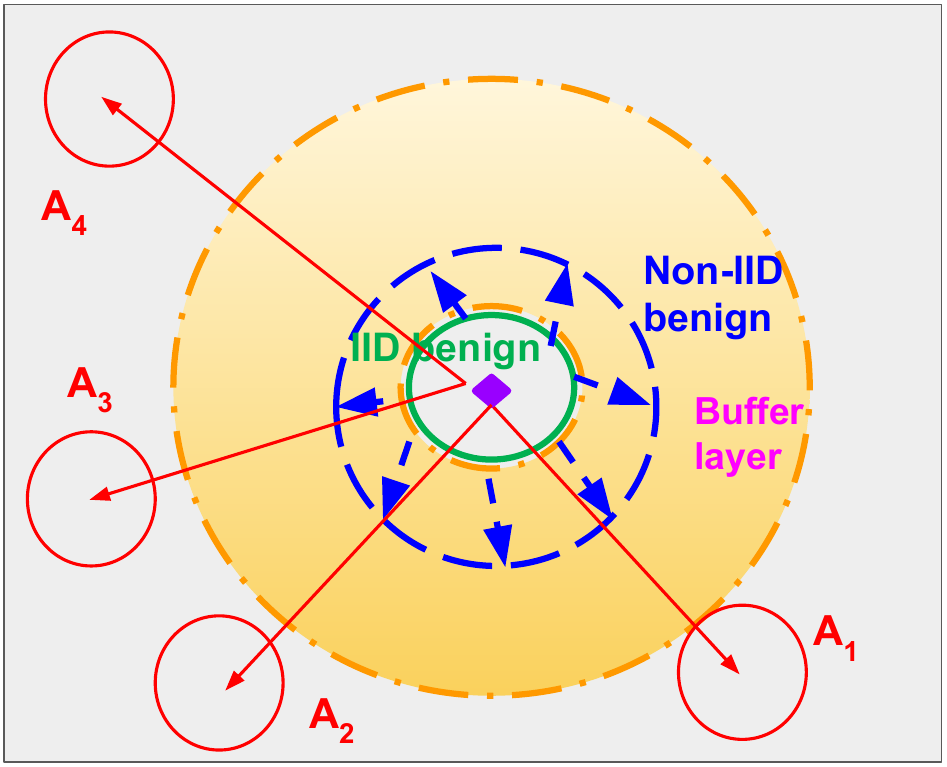}
        \caption{Modelling different backdoors}
        \label{fig:multi-attack-feasibility}
    \end{subfigure}
    \vspace{-5pt}
    \caption{A unified view for modeling the impacts of backdoor attacks and non-iids.}
    \label{fig:a-unified-view}
    \vspace{-5pt}
\end{figure*}

\subsection{Intuition} \label{sec:a-unified-view}
\label{subsec:design-uni-backdoor-niid}
Our intuition for \FB is based on our unique perspective for non-iids and backdoor attacks, illustrated in Fig.~\ref{fig:a-unified-view}. We can see that in representation space, the green solid circle, the blue dotted circle, and the red solid circle denote the cluster of iid benign model updates, that of non-iid benign model updates, and finally that of malicious model updates, respectively. On the one hand, we model \textit{the impacts of different non-iids as different expansions from the green circle to the blue one} under different sets of angles and displacements, uniquely determined by each individual \NIID. On the other hand, we model \textit{the impacts of one backdoor attack as a uni-directional displacement from the green circle to the red one}. What's more, defense performance degradation comes from that the blue circle starts to overlap with the red circle as a result of expansion (see Fig.~\ref{fig:small-buffer}). Therefore, \FB focuses on creating a sufficiently large buffer layer between the expanded blue circle and the red circle so that \FB can work under different \NIIDs (see Fig.~\ref{fig:large-buffer}). \FB is expected to work on different backdoor attacks as well because a larger buffer keeps all red circles (i.e., various backdoor attacks) out of the zone of the blue circle (see Fig.~\ref{fig:multi-attack-feasibility}).

\subsection{FLBuff: Buffer is All You Need}
\label{subsec:design-fb}
Based on the above intuition, the core design of \FB is to create a sufficiently large buffer layer in the representation space between the cluster of benign updates under \NIIDs and that of malicious updates. To that end, we design \FB as an aggregation-based defense which uses penultimate layer representations (PLRs) to denote model updates from remote clients and a supervised contrastive learning model to further train better representations from labeled PLRs. Fig.~\ref{fig:flbuff-flow} shows the workflow of \FB as explained below.

\textbf{Step 1:} A selected client $u_i$ needs to send her model update $\delta_i$ after local training to PS. 

\textbf{Step 2:} \FB is to compute the PLR sequence for each received $\delta_i$. Following the SOTA defenses~\cite{cao2021fltrust,wang2022flare,CrowdGuard}, we assume that PS has a small set of auxiliary data (denoted by $D_A$) for computing PLR sequences. We show later in Sect.~\ref{sec:evaluation:ablation} that $D_A$ can be samples from public generic dataset as well. Given $\delta_i$, {\FB first aggregates $\Theta$ and $\delta_i$ together and obtains $\Theta_i$.} The PLR sequence of $\delta_i$ can be readily obtained by feeding $D_A$ to $\Theta_i$ and then extracting the per-sample representations at the penultimate layer, i.e., the second last layer, and concatenating them one by one. For instance, assuming that the mapping function of $\Theta_i$ is $g_{\Theta_i}:x\in \mathcal{R}^{l \times w \times h} \rightarrow r \in \mathcal{R}^{p}$ where $l, w$, and $h$ denote the width, height, and channels of an image input, respectively, and $p$ the size of PLR. The PLR sequence of $\delta_i$ is thus $\text{PSeq}_i =: \{g_{\Theta_i}(x_1),\ldots,g_{\Theta_i}(x_m)\}$, where $m$ is the size of $D_A$.

\textbf{Step 3:} Here \FB is to train a supervised contrastive learning (Sup-CL) model so as to obtain better representations for benign and malicious PLRs. Given a set of labeled PLRs (i.e., benign and malicious), we train the Sup-CL model using the below loss function: 

\begin{equation}
    \text{loss}_{\text{Sup-CL}}=\frac{1}{\alpha(\alpha-1)} \sum_{i=1}^{\alpha}\sum_{j=i+1}^{\alpha} \left({L}_{i,j}+ L_{j,i}\right)\;,
    \label{eq:loss_cl}
\end{equation}
where 
\begin{equation}
\fontsize{6}{6}\selectfont
    L_{i,j}=\\
    -\log\frac{e^{\mathcal{G}(\text{PSeq}_{b,i})^{T}\mathcal{G}(\text{PSeq}_{b,j}) /\tau}}{e^{\mathcal{G}(\text{PSeq}_{b,i})^{T}\mathcal{G}(\text{PSeq}_{b,j}) /\tau}+\sum_{z=1}^{\beta} e^{\mathcal{G}(\text{PSeq}_{b,i})^{T}\mathcal{G}(\text{PSeq}_{m,z})/\tau}}\;,
    \label{eq:loss_cl_2}
\end{equation} 
and $\mathcal{G}(\cdot)$ is the encoder and $\tau\in[0,1]$ a parameter that controls distribution concentration level~\cite{TianWha20}. 

\textbf{Step 4:} \FB is to compute a trust score for each $\delta_i$ by first using Maximum Mean Discrepancy~\cite{ArthurMMD12} (MMD) to compute a distance $\text{MMD}(\text{PSeq}_i, \text{PSeq}_j)$ between two PLR sequences and then estimating the trust score of $\delta_i$ by counting the number of neighbors based on the computed distances: 


\begin{itemize}
    \item For $\delta_i$, \FB selects the top 50\% nearest neighbors based on the MMD distances in current round. The count for $\delta_i$, denoted by \( \text{Count}_i \), is to increase by one every time $\delta_i$ is selected as a nearest neighbor by another arbitrary \( \delta_j \) (\( j \neq i \)).
    \item \FB converts \( \text{Count}_i \) into a trust score \( \text{Score}_i \) using a softmax function: $\text{Score}_i = \frac{\exp(\text{Count}_i / \kappa)}{\sum_{k=1}^n \exp(\text{Count}_k / \kappa)}$,
    where \( \kappa \) is a temperature parameter to adjust the sensitivity of $\text{Score}_i$. \( \kappa = 1 \) is used in \FB.
\end{itemize}

\textbf{Step 5:} Finally, \FB aggregates all $\delta_i$s using $\text{Score}_i$ and updates $\Theta$ by $\Theta \leftarrow \Theta + \sum_{i=1}^n \text{Score}_i \cdot \delta_i.$




\section{Performance Evaluation}
\label{sec:evaluation}

\begin{table*}[ht]
\centering

\renewcommand{\arraystretch}{1.2} 
\caption{\texttt{ASR} (\%) of backdoor attacks with \FB and five baselines under iid and five non-iids.}
\label{tab:noniid_bechmark}
\resizebox{\textwidth}{!}{
\begin{tabular}{c|c|cccccc|cccccc|cccccc|cccccc}
\hline
\multirow{2}{*}{\textbf{Defense}} & \multirow{2}{*}{\textbf{Dataset}} & \multicolumn{6}{c|}{\textbf{BadNets}} & \multicolumn{6}{c|}{\textbf{DBA }} & \multicolumn{6}{c}{\textbf{Scaling }} & \multicolumn{6}{c}{\textbf{ALIE }} \\
\cline{3-26}
 &  & \textbf{iid} & \textbf{dir} & \textbf{prob} & \textbf{qty} & \textbf{noise} & \textbf{qs}  & \textbf{iid} & \textbf{dir} & \textbf{prob} & \textbf{qty} & \textbf{noise} & \textbf{qs} & \textbf{iid} & \textbf{dir} & \textbf{prob} & \textbf{qty} & \textbf{noise} & \textbf{qs} & \textbf{iid} & \textbf{dir} & \textbf{prob} & \textbf{qty} & \textbf{noise} & \textbf{qs}  \\
\hline
\multirow{3}{*}{\textbf{FLTrust}} 
    & MNIST & \default{0.90} & \default{2.01} & \default{2.70} & \default{4.99} & \default{1.70} & \default{8.16} & \default{0.78} & \default{2.25} & \default{3.29} & \default{6.12}  & \default{3.12} & \default{10.75} & \default{0.46} & \default{4.17} & \default{3.20} & \default{10.96}  & \default{2.40} & \default{9.71} & \default{0.84} & \default{4.51} & \default{5.02} & \default{6.61} & \default{2.42} & \default{10.85} \\
    & FMNIST & \default{0.63} & \default{4.07} & \default{3.03} & \default{7.36}  & \default{2.90} & \default{10.24} & \default{1.46} & \default{4.59} & \default{6.20} &  \default{8.50} & \default{3.20} & \default{9.57} & \default{2.80} & \default{3.87} & \default{5.38} & \default{11.88}  & \default{5.00} & \default{8.40} & \default{0.78} & \default{3.42} & \default{2.01} & \default{8.55} & \default{4.84} & \default{7.26} \\
    & CIFAR10 & \default{2.51} & \default{10.09} & \default{7.12} & \default{10.80}  & \default{5.72} & \default{13.44} & \default{1.25} & \default{8.71} & \default{7.73} & \default{13.39} & \default{8.33} & \default{10.62} & \default{0.89} & \default{13.21} & \default{7.70} & \default{15.31}  & \default{4.32} & \default{48.01} & \default{0.94} & \default{11.29} & \default{8.79} & \default{17.54} & \default{6.18} & \default{12.35} \\
    & IMDB & \default{3.09} & \default{10.22} & \default{13.17} & \default{10.75}  & \default{5.40} & \default{11.60} & \default{4.03} & \default{7.35} & \default{8.04} & \default{11.01}  & \default{4.50} & \default{12.57} & \default{3.27} & \default{7.09} & \default{10.55} & \default{9.15}  & \default{3.90} & \default{11.27} & \default{4.23} & \default{9.47} & \default{7.41} & \default{10.31} & \default{5.91} & \default{13.88} \\
\hline
\multirow{3}{*}{\textbf{FLARE}} 
    & MNIST & \default{0.48} & \default{3.10} & \default{2.77} & \default{2.40}  & \default{2.08} & \default{12.42} & \default{0.42} & \default{2.27} & \default{2.52} & \default{0.94}  & \default{3.44} & \default{8.14} & \default{0.52} & \default{1.25} & \default{2.92} & \default{5.56}  & \default{3.79} & \default{10.12} & \default{0.32} & \default{6.57} & \default{5.01} & \default{7.91} & \default{3.34} & \default{8.66} \\
    & FMNIST & \default{1.00} & \default{7.10} & \default{5.80} & \default{7.31}  & \default{5.97} & \default{14.43} & \default{4.40} & \default{5.90} & \default{6.92} & \default{4.25}  & \default{5.17} & \default{10.03} & \default{4.72} & \default{15.80} & \default{2.90} & \default{6.98}  &  \default{5.43} & \default{15.45} & \topone{0.10} & \default{7.31} & \default{8.19} & \default{10.05} & \default{3.30} & \default{12.91} \\
    & CIFAR10 & \default{1.80} & \default{10.81} & \default{6.90} & \default{9.77}  & \default{10.10} & \default{20.01} & \default{1.50} & \default{10.40} & \default{13.10} & \default{3.72}  &  \default{10.80} & \default{17.90} & \default{2.10} & \default{11.60} & \default{12.60} & \default{8.89}  & \default{10.97}  & \default{13.50} & \default{1.14} & \default{12.06} & \default{13.56} & \default{8.45} & \default{9.45} & \default{18.61} \\
    & IMDB & \default{2.47} & \default{10.06} & \default{6.17} & \default{11.57}  & \default{9.30} & \default{11.20} & \default{2.92} & \default{8.89} & \default{10.22} & \default{6.98}  & \default{12.30} & \default{14.40} & \default{1.91} & \default{7.54} & \default{8.18} & \default{8.95}  & \default{8.56} &  \default{12.30} & \default{2.81} & \default{5.52} & \default{10.20} & \default{11.13} & \default{9.90} & \default{12.10} \\
\hline
\multirow{3}{*}{\textbf{FLDetector}} 
    & MNIST & \default{0.26} & \default{8.80} & \default{5.96} & \default{10.54}  & \default{1.10} & \default{10.22} & \default{1.53} & \default{3.41} & \default{8.96} & \default{12.51}  & \default{1.57} & \default{6.39} & \default{2.01} & \default{4.60} & \default{4.19} & \default{16.52}  & \default{2.79} & \default{10.58} & \default{0.95} & \default{5.17} & \default{5.30} & \default{8.90} & \default{3.78} & \default{10.92} \\
    & FMNIST & \default{0.29} & \default{9.14} & \default{4.28} & \default{12.05}  & \default{2.70} & \default{15.25} & \default{3.89} & \default{4.13} & \default{6.51} & \default{15.43}  &  \default{5.00} & \default{11.10} & \default{2.89} & \default{7.00}  & \default{5.90} & \default{18.74}  & \default{3.60} & \default{15.34} & \default{1.87} & \default{5.52} & \default{4.13} & \default{12.78}  & \default{4.93} & \default{14.10} \\
    & CIFAR10 & \default{3.26} & \default{8.42} & \default{8.31} & \default{16.50}  & \default{10.10} & \default{14.90} & \default{3.43} & \default{10.57} & \default{8.24} & \default{15.50}  & \default{8.70} & \default{22.60} & \default{6.60} & \default{7.08} & \default{5.10} & \default{16.95}  & \default{10.02}  & \default{27.88} & \default{1.93} & \default{9.29} & \default{5.41} & \default{10.93} & \default{8.50} & \default{35.80} \\
    & IMDB & \default{3.15} & \default{6.10} & \default{5.30} & \default{19.25}  & \default{9.68} & \default{16.54} & \default{2.82} & \default{7.52} & \default{5.46} & \default{18.83}  &  \default{5.72} & \default{25.52} & \default{3.02} & \default{5.46} & \default{5.02} & \default{7.51}  & \default{8.10} & \default{10.90} & \default{4.70} & \default{7.35} & \default{6.23} & \default{8.50} & \default{5.40} & \default{10.90} \\
\hline
\multirow{3}{*}{\textbf{FLAME}} 
    & MNIST & \topone{0.21} & \default{1.35} & \default{3.37} & \default{3.32}  & \default{0.63} & \default{2.29} & \topone{0.30} & \default{1.46} & \default{1.88} & \default{1.48}  & \topone{0.31} & \default{3.96} & \default{0.44} & \default{1.71} & \default{5.18} & \default{1.64}  & \default{1.88} & \default{5.11} & \topone{0.28} & \default{1.52} & \default{2.82} & \default{3.85} & \topone{0.94} & \default{5.66} \\
    & FMNIST & \default{0.75} & \default{2.10} & \default{6.58} & \default{7.86}  & \default{1.52} & \default{8.10} & \default{0.43} & \default{3.30} & \default{3.55} & \default{4.53}  & \default{3.64} & \default{7.24} & \default{0.91} & \default{5.51} & \default{8.23} & \default{6.15}  & \default{4.37} & \default{6.10} & \default{1.74} & \default{5.70} & \default{6.72} & \default{5.60} & \default{2.10} & \default{8.00} \\
    & CIFAR10 & \default{3.20} & \default{5.10} & \default{6.50} & \default{11.10}  & \default{7.80} & \default{18.60} & \default{3.30} & \default{10.60} & \default{7.70} & \default{14.51}  &  \default{6.50} & \default{11.80} & \default{1.80} & \default{10.60} & \default{6.80} & \default{10.72}  & \default{4.10} &  \default{11.00} & \default{1.15} & \default{7.80} & \default{10.70} & \default{11.25} & \default{2.73} &  \default{14.50} \\
    & IMDB & \default{1.02} & \default{3.35} & \default{7.15} & \default{9.90}  & \default{5.80} & \default{10.00} & \default{1.63} & \default{6.45} & \default{10.68} & \default{9.95}  & \default{4.83} & \default{9.64} & \default{2.07} & \default{8.55} & \default{6.73} & \default{10.16}  & \default{6.10} & \default{10.10} & \default{2.12} & \default{8.75} & \default{6.85} & \default{10.30} & \default{3.00} & \default{10.60} \\
\hline
\multirow{3}{*}{\textbf{FLIP}} 
    & MNIST & \default{1.13} & \default{5.11} & \default{7.27} & \default{8.32}  & \default{1.42} & \default{4.87} & \default{2.10} & \default{10.82} & \default{11.40} & \default{12.38}  & \default{5.35} & \default{2.44} & \default{1.22} & \default{5.53} & \default{8.98} & \default{9.54}  & \default{2.10} &  \default{4.68} & \default{3.21} & \default{8.65} & \default{12.52} & \default{8.75} & \default{3.85} & \default{7.51} \\
    & FMNIST & \topone{0.30} & \default{4.98} & \default{4.58} & \default{10.76}  & \default{2.71} & \default{5.38} & \default{4.14} & \default{10.41} & \default{8.85} & \default{12.63}  & \default{1.24} & \default{6.90} & \default{1.11} & \default{12.36} & \default{9.13} & \default{9.05}  & \default{0.73}  &  \default{5.97} & \default{2.14} & \default{7.36} & \default{8.82} & \default{8.70} & \default{4.90} & \default{4.83} \\
    & CIFAR10 & \default{3.20} & \default{8.78} & \default{9.90} & \default{14.30}  & \default{6.88} & \default{24.20} & \default{6.60} & \default{7.44} & \default{8.80} & \default{10.61}  & \default{12.70} & \default{27.90} & \default{5.60} & \default{8.69} & \default{7.90} & \default{10.81}  & \default{7.20} & \default{15.40} & \default{1.45} & \default{10.91} & \default{11.10} & \default{14.35} & \default{9.00} & \default{21.80} \\
    & IMDB & \default{3.12} & \default{9.45} & \default{10.25} & \default{12.00}  & \default{4.50} & \default{10.20} & \default{5.73} & \default{13.55} & \default{7.78} & \default{12.05}  & \default{5.45} & \default{11.70} & \default{2.17} & \default{21.65} & \default{22.83} & \default{13.26}  & \default{3.70} & \default{8.40} & \default{3.22} & \default{11.85} & \default{9.95} & \default{13.40} & \default{6.20} & \default{15.30} \\
\hline
\multirow{3}{*}{\textbf{FLBuff}} 
    & MNIST & \default{0.30} & \topone{1.00} & \topone{0.70} & \topone{0.52}  & \topone{0.98} & \topone{0.65} & \default{0.56} & \topone{1.27} & \topone{0.90} & \topone{1.40}  & \default{0.70} & \topone{0.85}  & \topone{0.15} & \topone{0.12} & \topone{0.43} & \topone{0.28}  & \topone{0.71} & \topone{0.60} & \default{0.93} & \topone{1.07} & \topone{1.25} & \topone{0.42} & \default{1.14} & \topone{0.53} \\
    & FMNIST & \default{0.85} & \topone{0.53} & \topone{0.87} & \topone{1.23}  & \topone{0.53} & \topone{0.51} & \topone{0.15} & \topone{0.25} & \topone{0.47} & \topone{0.22}  & \topone{1.12} & \topone{0.49} & \topone{0.62} & \topone{0.31} & \topone{0.92} & \topone{0.76}  & \topone{0.65} & \topone{0.81} & \default{0.76} & \topone{0.74} & \topone{0.38} & \topone{1.07} & \topone{0.95} & \topone{0.85}  \\
    & CIFAR10 & \topone{0.30} & \topone{0.93} & \topone{0.94} & \topone{0.77}  & \topone{0.80} & \topone{0.62}  & \topone{0.51} & \topone{1.28} & \topone{0.55} & \topone{1.42}  & \topone{1.19}  & \topone{0.65} & \topone{0.35} & \topone{1.13} & \topone{0.77} & \topone{0.81}  & \topone{0.54} & \topone{1.20} & \topone{0.04} & \topone{0.19} & \topone{1.26} & \topone{1.01} & \topone{0.68} & \topone{1.09} \\
    & IMDB & \topone{0.35} & \topone{0.33} & \topone{0.89} & \topone{0.50} & \topone{0.60} & \topone{0.94} & \topone{0.62} & \topone{1.18} & \topone{1.10} & \topone{1.18}  & \topone{0.74} & \topone{0.70} & \topone{0.85} & \topone{1.26} & \topone{0.93} & \topone{1.05} & \topone{0.72} & \topone{1.05} & \topone{0.97} & \topone{0.22} & \topone{0.29} & \topone{0.86} & \topone{0.84} & \topone{0.91} \\
\hline

\end{tabular}
}
\end{table*}

\subsection{Experimental Settings \& Goals}
\label{eval:setup}

\subsubsection{Implementation} 
We implemented \FB in PyTorch. The experiments were executed on a server with two NVIDIA RTX 3090 GPUs. 

\subsubsection{Datasets and Models} 
\label{non-iid-datasets}

We evaluated \FB on three image datasets as in~\cite{wang2022flare,zhangfldetecotr,CrowdGuard} and one additional text dataset: MNIST, Fashion-MNIST (FMNIST), CIFAR10, and IMDB. For iid scenarios, we divide the data equally among clients. For non-iids, we consider five \NIID types with the following default non-iid parameters: \textbf{1)} prob-based (0.3), \textbf{2)} dir-based (0.3),  \textbf{3)} qty-based where each client only has two classes, \textbf{4)} noise-based ($\sigma=0.5$), and \textbf{5)} quantity skew with $\beta=0.5$ for Dirichlet distribution. We also explore different \NIID degrees in ablation study at Sect.~\ref{noniid-degree-study}.

We adopt a CNN model for MNIST and FMNIST, ResNet-50 for CIFAR10, and LSTM for IMDB. The details of these models are in \textbf{supplementary}. 


    

\subsubsection{FL Settings} 
In each FL round, we select 10 out of 100 clients via \textbf{random selection} to participate FL training. By default, the malicious client ratio is set to 0.2 and the data poisoning rate 0.5. Each client trains its local model using a batch size of 64 for five local epochs. {The classification threshold is set to 0.5 for the Sup-CL module.} For MNIST and FMNIST, we adopt cross-entropy as the loss function, stochastic gradient descent (SGD) with a learning rate of 0.001, and a momentum of 0.9 to train models. For CIFAR10, we use the same loss function, optimizer and momentum decade but adopt a learning rate of 0.01 for SGD instead. For IMDB, we adopt cross-entropy with logit as the loss function and Adam optimizer with a learning rate of 0.001.


\subsubsection{Metrics}
We use both attack success rate (\texttt{ASR}) and model accuracy (\texttt{Acc}) to evaluate FLBuff as in~\cite{wang2022flare,zhangfldetecotr,CrowdGuard}. \texttt{ASR} denotes the ratio of backdoor samples that successfully achieve targeted predictions from the victim FL model. The lower \texttt{ASR}, the better \FB performs. \texttt{Acc} denotes the classification accuracy of the targeted FL model (i.e., the victim model) during testing.

\subsubsection{Baselines} 
\label{sec:attacks-and-defenses-used}
We evaluate \FB against four backdoors: \textit{BadNets}~\cite{gu2017badnets}, \textit{DBA}~\cite{xie2019dba}, \textit{Scaling}~\cite{bagdasaryan2020backdoor}, and \textit{ALIE}~\cite{baruch2019little}. We compare \FB to five SOTA defenses: \textit{FLTrust}~\cite{cao2021fltrust}, \textit{FLDetector}~\cite{zhangfldetecotr}, \textit{FLARE}~\cite{wang2022flare}, \textit{FLAME}~\cite{nguyen2022flame}, and \textit{FLIP}~\cite{zhang2023flipprovabledefenseframework}.
 
\subsubsection{Evaluation Goals}
{
We introduce our evaluation goals here:  
\textbf{Goal 1:} providing comprehensive attack and defense benchmarks under multiple \NIID settings, \textbf{Goal 2:} evaluating the effectiveness of FLBuff, and finally \textbf{Goal 3:} evaluating defense effectiveness under various settings including an adaptive attacker. }

\subsection{Experimental Results}
\subsubsection{Performance of Global Models}
Table~\ref{tab:acc_all_benigh} shows the accuracy of \textbf{clean global models without attacks} under iid and five \NIIDs. The results indicate that model accuracy decreases in \NIIDs due to \NIIDs. It also shows that \FB does not negatively impact the accuracy of the global model.

\begin{table}[!ht]
    \renewcommand{\arraystretch}{1.3}
    \centering
    \footnotesize
    \caption{\texttt{Acc} without backdoor attacks on FL.}
    \label{tab:acc_all_benigh}
    \begin{tabular}{p{1.7cm}|p{0.6cm}p{0.6cm}p{0.6cm}p{0.6cm}p{0.6cm}p{0.6cm}}
        \hline
         \centering
      \textbf{Dataset} & \textbf{iid} & \textbf{dir} & \textbf{prob} & \textbf{qty} & \textbf{noise} & \textbf{qs} \\ 
        \hline
\centering
     MNIST & \textbf{99.50} & 95.84 & 94.23 &  94.18 & 97.36 & 94.27 \\
     \centering
FMNIST & \textbf{85.51} & 80.86 & 79.68 & 79.30 & 80.58 & 80.10 \\
\centering
CIFAR-10 & \textbf{68.53} & 65.49 & 65.20 & 65.59 &  67.50 & 65.14 \\
\centering
IMDB & \textbf{87.60} & 85.20 & 83.36 & 82.52 & 85.69 & 84.01 \\
        \hline
    \end{tabular}
    \vspace{-5pt}
\end{table}

\subsubsection{Performance of FLBuff and Non-iid Benchmark}
\label{sec:evaluation:flbuff}
Table~\ref{tab:noniid_bechmark} shows the \texttt{ASR} of \FB and five baselines against four backdoor attacks under iid and five \NIIDs. \textbf{Note that} the results of \FB are obtained when applying the same \FB model trained from labeled benign and BadNets samples only to all backdoors (i.e., BadNets, DBA, Scaling, and ALIE) and all non-iids (i.e., dir, prob, qty, noise, and qs). Below are our findings from the results. \textbf{(1)} Existing defenses are not as robust as claimed when evaluated under different \NIIDs. For most defenses, their performance drops significantly when the scenario goes from iid to an arbitrary \NIID. For instance, \texttt{ASR} of FLTrust under \NIIDs changes vastly between 0.46\% to 48.01\%. {\textbf{(2)} While certain defenses may remain effective under a specific \NIID type, it is rare that they perform well across more than two \NIIDs}. One possible reason is the lack of built-in design for addressing \NIIDs in these defenses. \textbf{(3)} In comparison, \FB achieves the best performance among all. Most \texttt{ASR} using the same \FB model across various non-iids and backdoors is low around 1\% while \texttt{ASR} with other defenses easily exceeds 15\% or even 20\%. The results in Table~\ref{tab:noniid_bechmark} confirm \FB can consistently defend FL models against various backdoors under different non-iids.

We want to emphasize that our work provides a friendly and extendable benchmark for any FL evaluations regarding non-iids, which has been missing in the literature.

\subsubsection{Ablation Study}
\label{sec:evaluation:ablation}

\begin{figure*}[ht]
    \centering
    \begin{subfigure}[b]{0.24\textwidth}
        \includegraphics[width=\textwidth]{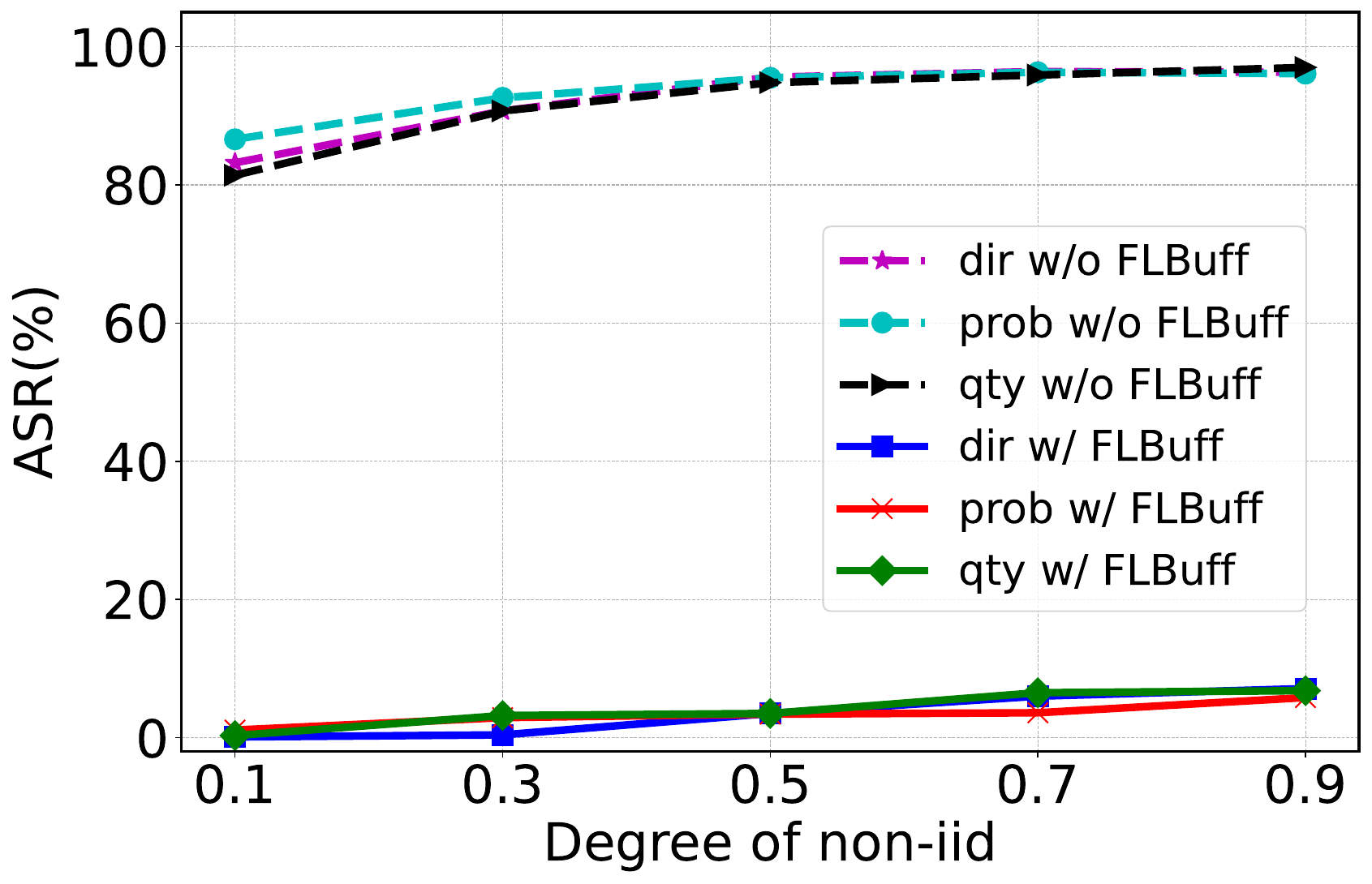}
        \caption{non-iid degree}
        \label{fig:noniid_degree_Scaling}
    \end{subfigure}
    \begin{subfigure}[b]{0.24\textwidth}
        \includegraphics[width=\textwidth]{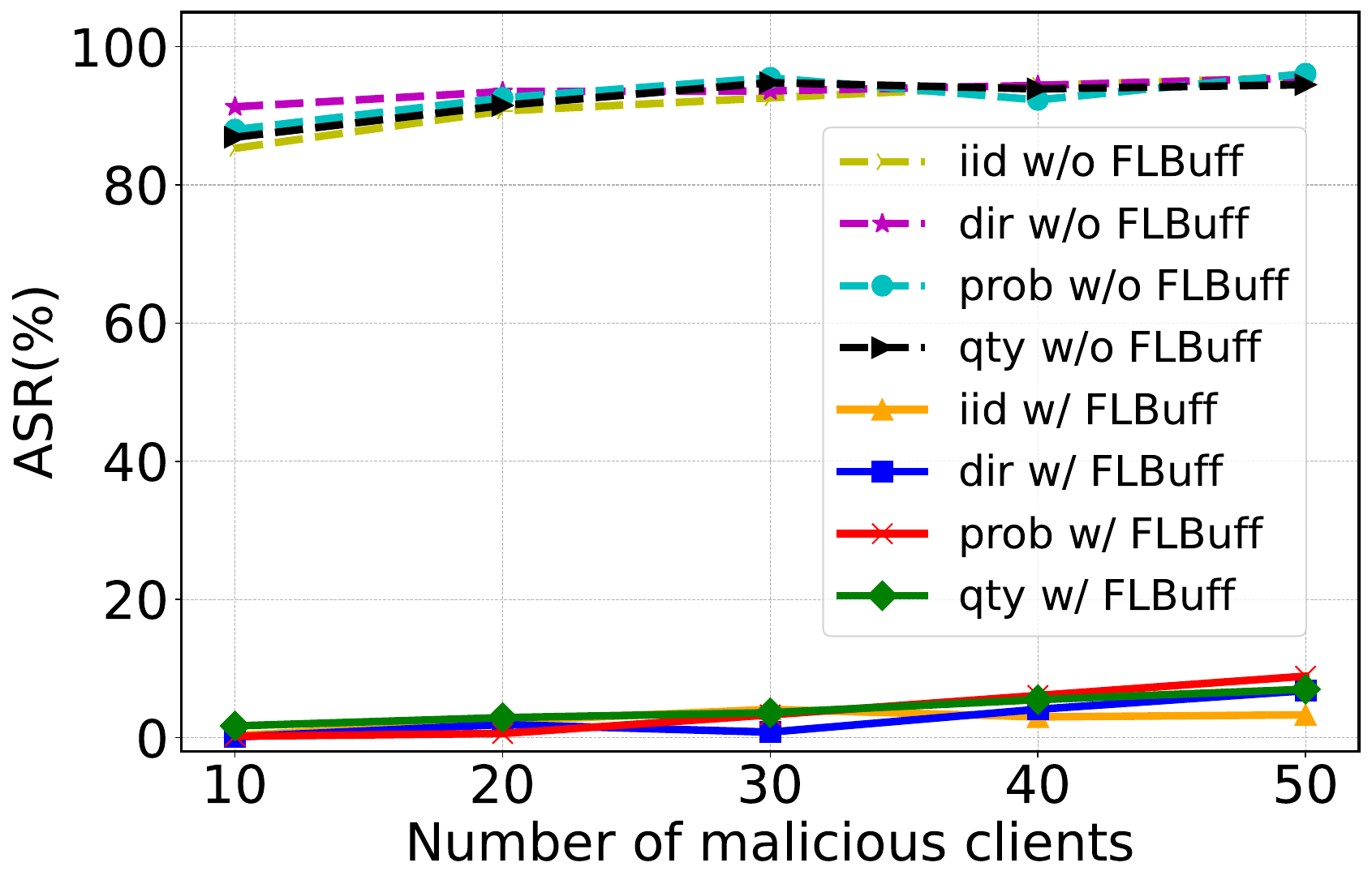}
        \caption{Number of attackers}
        \label{fig:rate_attackers_Scaling}
    \end{subfigure}
    \begin{subfigure}[b]{0.24\textwidth}
        \includegraphics[width=\textwidth]{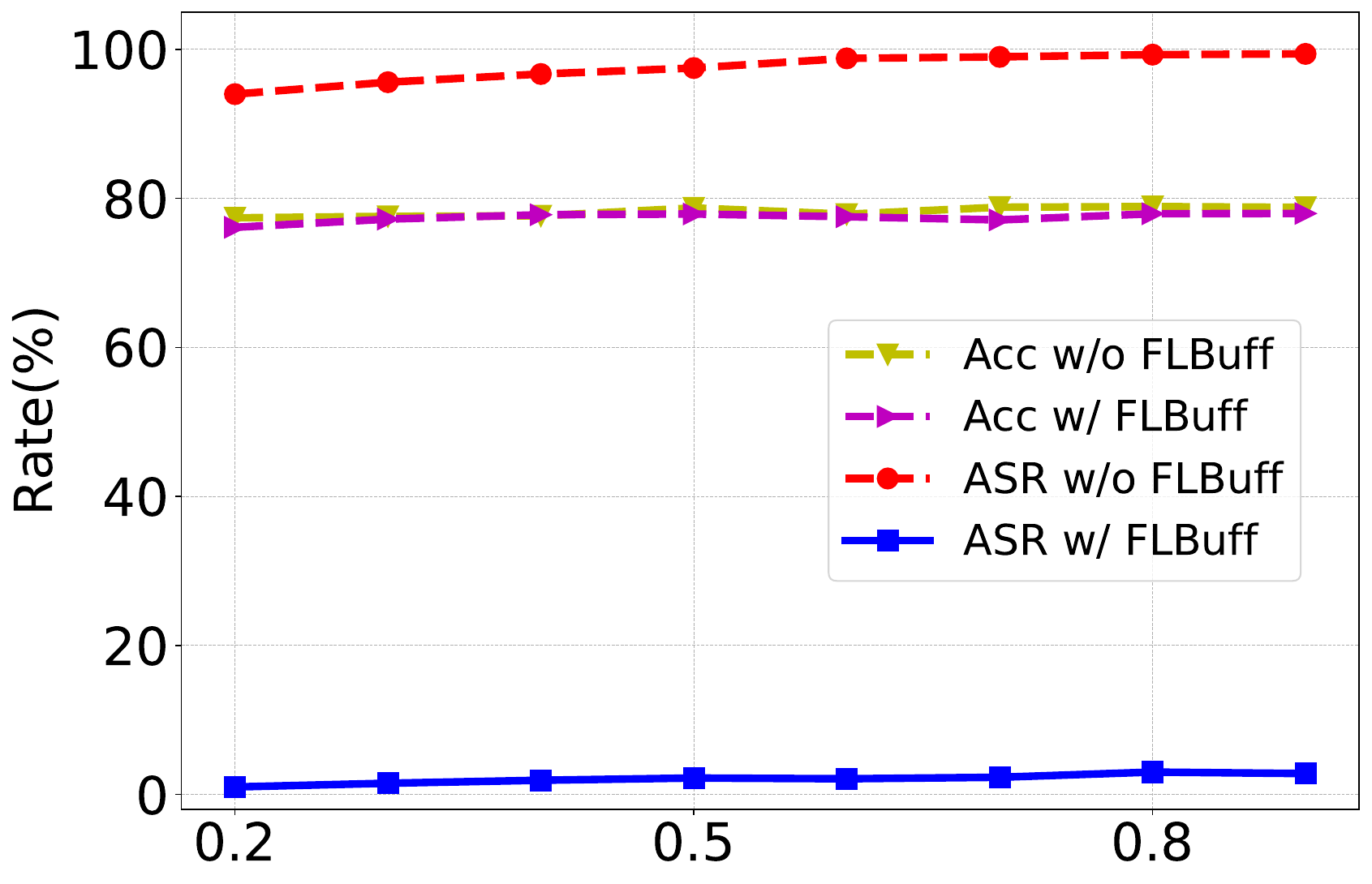}
        \caption{Data poisoning rate}
        \label{fig:data_poisoning_rate}
    \end{subfigure}
    \begin{subfigure}[b]{0.24\textwidth}
        \includegraphics[width=\textwidth]{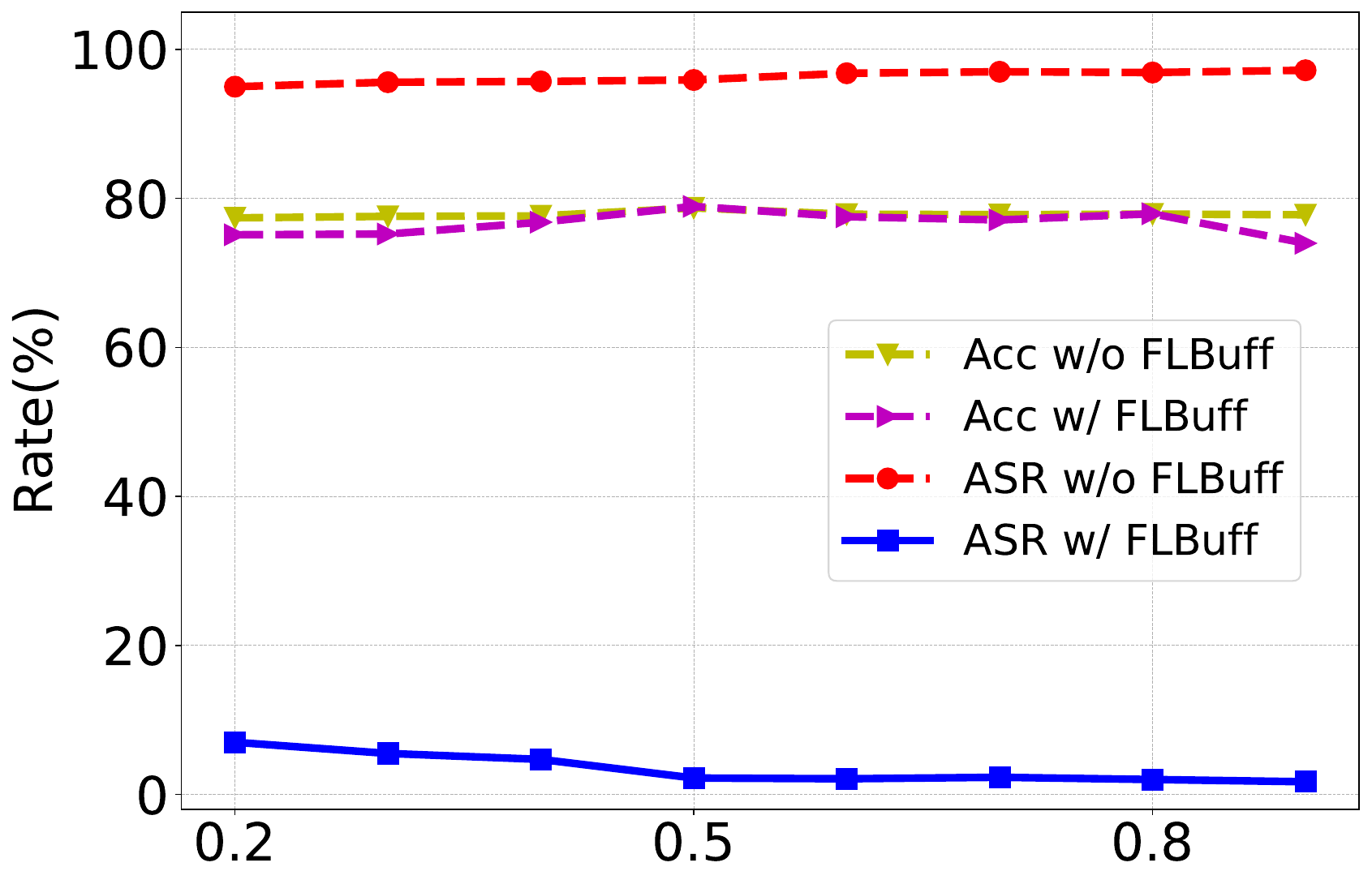}
        \caption{Classification threshold}
        \label{fig:detection_threshold}
    \end{subfigure}
    \vspace{-5pt}
    \caption{Impact of various factors on FLBuff on CIFAR10 with Scaling attack.}
    \label{fig:noniid-degree-impact}
    \vspace{-5pt}
\end{figure*}


\textbf{Impact of non-iid degree.}
\label{noniid-degree-study}
For convenience, we first normalize all non-iid degrees to $[0,1]$. Intuitively, a large \NIID degree leads to severe expansion of \NIID, which makes it more difficult to defend (see Fig.~\ref{fig:small-buffer}). Fig.~\ref{fig:noniid_degree_Scaling} shows the impact of \NIID degree on \FB, from which we have two observations. \textbf{(1)} The \texttt{ASR} without \FB increases as \NIID degree increases indicating that a larger \NIID degree harms the FL model more severely. \textbf{(2)} In comparison, when \NIID degree increases, it only impacts \FB slightly, hence rendering \FB a robust defense towards \NIID degree. {Results for `noise' and `qs' can be found in \textbf{supplementary}.}


\textbf{Impact of the number of attackers.}
Fig.~\ref{fig:rate_attackers_Scaling} shows the impact of the number of malicious clients on \FB. \textbf{(1)} We observe that without \FB, more attackers can lead to higher \texttt{ASR} under iid and \NIID settings. \textbf{(2)} \FB can effectively mitigate backdoors as it achieves low \texttt{ASR} even when the number of attackers increases. 


\textbf{Impact of data poisoning rate.} 
Fig.~\ref{fig:data_poisoning_rate} shows \texttt{ASR} at different poisoning rates. \textbf{(1)} \texttt{ASR} increases with higher poisoning rates without \FB. \textbf{(2)} \FB remains largely unaffected with increasing rates.

\textbf{Impact of classification threshold $\gamma$.} 
Fig.~\ref{fig:detection_threshold} illustrates the impact of classification threshold on \FB. The results confirm that $\gamma$ would not affect \texttt{ASR} of \FB much as long as it is not too low (e.g., $\le 0.2$) nor too high (e.g., $\ge 0.9$). 

\textbf{Impact of auxiliary dataset.}
We tested if public datasets could replace the original auxiliary data in Step 2 as alternatives, e.g., using EMNIST samples for MNIST. Table~\ref{tab:other-auxiliary} shows that this led to minimal changes in \texttt{ASR}, confirming that \FB would not rely on specific auxiliary datasets.

\begin{table}[!ht]
    \renewcommand{\arraystretch}{1.0}
    \centering
    \caption{Impact of auxiliary datasets on \FB (\texttt{ASR}), Scaling}
    \scriptsize 
    \resizebox{\columnwidth}{!}{ 
    \begin{tabular}{p{0.8cm}|p{1.45cm}|p{0.5cm}p{0.5cm}p{0.5cm}p{0.5cm}p{0.5cm}p{0.5cm}}
    \hline
    \textbf{\#Seq} & \textbf{Train-Set \& Auxiliary-Set} & \textbf{iid} & \textbf{dir} & \textbf{prob} & \textbf{qty} & \textbf{noise} & \textbf{qs} \\ 
    \hline
    \multirow{2}{*}{Pair1} & MNIST & 0.20 & 0.10 & 0.50 & 0.30 & 0.70 & 0.60 \\
                           & EMNIST & 0.27 & 0.15 & 0.53 & 0.26 & 0.74 & 0.58 \\
    \hline
    \multirow{2}{*}{Pair2} & FMNIST & 0.69 & 0.40& 0.98 & 0.75 & 0.65 & 0.80 \\
                           & KMNIST & 0.75 & 0.50 & 1.10 & 0.90 & 0.70 & 0.86 \\
    \hline
    \multirow{2}{*}{Pair3} & CIFAR-10 & 0.40 & 1.10 & 0.72 & 0.80 & 0.55 & 1.22 \\
                           & CIFAR100  & 0.50 & 1.25 & 0.88 & 0.95  &  0.62  & 1.31 \\
    \hline
    \multirow{2}{*}{Pair4} & IMDB & 0.90 & 1.30 & 0.90 & 1.03 & 0.70  & 1.00 \\
                           & Sentiment-140 & 1.02 & 1.40 & 1.10 & 1.15 & 0.90 & 1.11 \\
    \hline
    \end{tabular}
    }
    \label{tab:other-auxiliary}
    \vspace{-5pt}
\end{table}

\begin{table}[!ht]
\centering
\renewcommand{\arraystretch}{1.0}
\caption{\texttt{ASR} (\%) with FLBuff under adaptive attacks}
\label{tab:adaptive_attack_table}
\resizebox{\columnwidth}{!}{
\begin{tabular}{p{1.7cm}|p{0.6cm}p{0.6cm}p{0.6cm}p{0.6cm}p{0.6cm}p{0.6cm}}
\hline
\centering
\textbf{Dataset} & \textbf{iid} & \textbf{dir} & \textbf{prob} & \textbf{qty} & \textbf{noise} & \textbf{qs} \\
\hline
\centering
MNIST    & \textbf{0.20} & 0.42 & 1.37 & 0.35 & 0.72 & 0.93 \\
\centering
FMNIST   & 0.96 & 1.35 & 1.86 & \textbf{0.60} & 1.38 & 1.50 \\
\centering
CIFAR-10 & 1.15 & 1.50 & 1.09 & \textbf{1.06} & 0.73 & 1.09 \\
\centering
IMDB     & 0.57 & 1.38 & 1.03 & \textbf{0.42} & 1.00 & 1.34 \\
\hline
\end{tabular}
}
\vspace{-5pt}
\end{table}

\subsubsection{Resilience to Adaptive Attacks under Non-iids} 


As in prior defenses, we expect that resilience to adaptive attacks is critical for \FB. Given knowledge about \FB, an attacker's goal is to use adaptive attacks to bypass \FB and successfully compromise $\Theta$. Our key differences from others are (1) prior works rarely evaluated a defense against adaptive attacks under \NIIDs while we do. Moreover, we believe \NIID settings are necessary for evaluating adaptive attacks because \NIIDs are fundamental in FL and effectively facilitate adaptive attacks. (2) We further consider stronger adaptive backdoors by normalizing the loss terms when formulating adaptive attacks. 

Our adaptive attacker is to craft her malicious model updates and the corresponding PLRs so that they resemble those of benign model updates. The formulation is as below while more details are in \textbf{supplementary.} 


\begin{equation}
    \begin{aligned}
        \arg\min_{\delta_{mal}} \quad L(\mathcal{D}_{mal}) + \lambda L(\mathcal{D}_{train}) + \rho \Tilde{\Delta d_{\delta}} + \eta \Tilde{d_{plr}},
    \end{aligned}
    \label{eq:adaptive-attack}
\end{equation}

\noindent where $L(\mathcal{D}_{mal})$ denotes the loss on targeted backdoored inputs, $L(\mathcal{D}_{train})$ the loss on clean dataset, $\Tilde{\Delta d_{\delta}} = \frac{\lVert \delta_{mal}-\bar{\delta}_{ben}\rVert}{\lVert \delta_{mal}-\bar{\delta}_{ben}\rVert_{max}}$ the normalized distance between malicious model updates and average benign ones, and finally \(\Tilde{d_{plr}}=\frac{d_{plr}}{d_{plr,max}}\) the normalized distance between the PLRs of malicious model updates and the average one of benign models. $\lambda$, $\rho$, and $\eta$ are the corresponding coefficients. 
Table~\ref{tab:adaptive_attack_table} lists the resulted \texttt{ASR} of the above adaptive attack on \FB. The results confirm that \FB still achieves low \texttt{ASR} against adaptive attacks particularly across various datasets and all the five \NIIDs.


\section{Related Work}
\label{sec:related}
\subsection{Backdoor Attacks}
The recent years have seen many novel backdoor attacks featuring their unique attack strategies. For instance,
BadNets~\cite{gu2017badnets} introduced an embedded trigger to enforce specific model outputs, while Xie et al.\cite{xie2019dba} extended this approach to distributed settings with DBA. Scaling\cite{bagdasaryan2020backdoor} and ALIE~\cite{baruch2019little} further demonstrated the feasibility of stealthy backdoor attacks. Our paper adopts recent popular backdoor attacks which have been shown to bypass strong defenses.

\subsection{Defenses against Backdoor Attacks}

FLTrust~\cite{cao2021fltrust} aggregated model updates by computing a trust score based on their direction and magnitude similarity. FLARE~\cite{wang2022flare} collected PLRs for better characterizing model updates and then computing trust scores for aggregation as well. FLAME~\cite{nguyen2022flame} used clustering and weight clipping to estimate and inject minimal noise for mitigating backdoors. Zhang et al.~\cite{zhang2023flipprovabledefenseframework} utilized reverse engineering to identify backdoor triggers. 
FLDetector~\cite{zhangfldetecotr} filtered out malicious model updates using historical data and consistency check. CrowdGuard~\cite{CrowdGuard}, MESAS~\cite{MESAS}, and FreqFed~\cite{fereidooni2023freqfed} are the few most recent defenses aiming to address \NIID challenges as well.  Aside from unique assumptions on hardware (e.g., CrowdGuard required TEE for secure computation) and backdoor strategies (e.g., FreqFed assumed unique attack patterns in frequency domain), they still suffer from insufficient non-iid comprehensiveness as shown in Table~\ref{table:flbuff-position}.






\textbf{Differences in FLBuff.} Firstly, \FB considered the most comprehensive and important set of \NIIDs. Secondly, the design of \FB started from the very root of why prior works fail under various \NIIDs. Thirdly, \FB considered stronger adaptive attacks under \NIIDs as well. 

\section{Conclusions}
\label{sec:conclude}
In this paper, we introduce \FB, a novel defense framework designed to protect FL against SOTA backdoor attacks under five types of challenging \NIIDs. \FB aims to fill the gap that existing defenses simply fail under different \NIIDs. We designed \FB as an aggregation-based defense aiming at creating a large buffer layer to separate benign and malicious clusters. Our extensive results demonstrate \FB's better performance in different scenarios, including against previously unseen attacks and strong adaptive attacks as well. 


{
\small
\bibliographystyle{ieeetr}
\bibliography{paper}

\begin{thebibliography}{10}

\bibitem{fang2020local}
M.~Fang, X.~Cao, J.~Jia, and N.~Gong, ``Local model poisoning attacks to byzantine-robust federated learning,'' in {\em USENIX Security}, 2020.

\bibitem{bagdasaryan2019differential}
E.~Bagdasaryan, O.~Poursaeed, and V.~Shmatikov, ``Differential privacy has disparate impact on model accuracy,'' in {\em NeurIPS}, 2019.

\bibitem{bagdasaryan2020backdoor}
E.~Bagdasaryan, A.~Veit, Y.~Hua, D.~Estrin, and V.~Shmatikov, ``How to backdoor federated learning,'' in {\em AISTATS}, 2020.

\bibitem{baruch2019little}
G.~Baruch, M.~Baruch, and Y.~Goldberg, ``A little is enough: Circumventing defenses for distributed learning,'' {\em NeurIPS}, 2019.

\bibitem{cao2021fltrust}
X.~Cao, M.~Fang, J.~Liu, and N.~Z. Gong, ``Fltrust: Byzantine-robust federated learning via trust bootstrapping,'' {\em NDSS}, 2021.

\bibitem{wang2022flare}
N.~Wang, Y.~Xiao, Y.~Chen, Y.~Hu, W.~Lou, and Y.~T. Hou, ``Flare: defending federated learning against model poisoning attacks via latent space representations,'' in {\em AsiaCCS}, 2022.

\bibitem{zhangfldetecotr}
Z.~Zhang, X.~Cao, J.~Jia, and N.~Z. Gong, ``Fldetector: Defending federated learning against model poisoning attacks via detecting malicious clients,'' in KDD, 2022.

\bibitem{zhang2023flipprovabledefenseframework}
K.~Zhang, G.~Tao, Q.~Xu, S.~Cheng, S.~An, Y.~Liu, S.~Feng, G.~Shen, P.-Y. Chen, S.~Ma, and X.~Zhang, ``Flip: A provable defense framework for backdoor mitigation in federated learning,'' 2023.

\bibitem{CrowdGuard}
P.~Rieger, T.~Krauß, M.~Miettinen, A.~Dmitrienko, and A.-R. Sadeghi, ``Crowdguard: Federated backdoor detection in federated learning,'' 2023.

\bibitem{MESAS}
T.~Krau\ss{} and A.~Dmitrienko, ``Mesas: Poisoning defense for federated learning resilient against adaptive attackers,'' in CCS, 2023.

\bibitem{fereidooni2023freqfed}
H.~Fereidooni, A.~Pegoraro, P.~Rieger, A.~Dmitrienko, and A.-R. Sadeghi, ``Freqfed: A frequency analysis-based approach for mitigating poisoning attacks in federated learning,'' {\em arXiv preprint arXiv:2312.04432}, 2023.

\bibitem{nguyen2022flame}
T.~D. Nguyen, P.~Rieger, R.~De~Viti, H.~Chen, B.~B. Brandenburg, H.~Yalame, H.~M{\"o}llering, H.~Fereidooni, S.~Marchal, M.~Miettinen, {\em et~al.}, ``$\{$FLAME$\}$: Taming backdoors in federated learning,'' in {\em USENIX Security}, 2022.

\bibitem{mcmahan2017communication}
B.~McMahan, E.~Moore, D.~Ramage, S.~Hampson, and B.~A. y~Arcas, ``Communication-efficient learning of deep networks from decentralized data,'' in {\em AISTATS}, 2017.

\bibitem{li2022federated}
Q.~Li, Y.~Diao, Q.~Chen, and B.~He, ``Federated learning on non-iid data silos: An experimental study,'' in {\em ICDE}, 2022.

\bibitem{luo2021no}
M.~Luo, F.~Chen, D.~Hu, Y.~Zhang, J.~Liang, and J.~Feng, ``No fear of heterogeneity: Classifier calibration for federated learning with non-iid data,'' {\em NeurIPS}, 2021.

\bibitem{xie2019dba}
C.~Xie, K.~Huang, P.-Y. Chen, and B.~Li, ``Dba: Distributed backdoor attacks against federated learning,'' in {\em ICLR}, 2019.

\bibitem{TianWha20}
Y.~Tian, C.~Sun, B.~Poole, D.~Krishnan, C.~Schmid, and P.~Isola, ``What makes for good views for contrastive learning?,'' in {\em NeurIPS}, 2020.

\bibitem{ArthurMMD12}
A.~Gretton, K.~M. Borgwardt, M.~J. Rasch, B.~Sch{{\"o}}lkopf, and A.~Smola, ``A kernel two-sample test,'' {\em Journal of Machine Learning Research}, vol.~13, no.~25, pp.~723--773, 2012.

\bibitem{gu2017badnets}
T.~Gu, B.~Dolan-Gavitt, and S.~Garg, ``Badnets: Identifying vulnerabilities in the machine learning model supply chain,'' {\em arXiv preprint arXiv:1708.06733}, 2017.

\end{thebibliography}
}

\end{document}